# Ultra-low Thermal Conductivity of Isotope Doped Silicon Nanowires

Nuo Yang[1], Gang Zhang[2*] and Baowen Li[1,3]

[1]*Department of Physics and Centre for Computational Science and Engineering, National University of Singapore, 117542 Singapore*

[2]*Institute of Microelectronics, 11 Science Park Road, Singapore Science Park II, Singapore 117685, Singapore*

[3]*NUS Graduate School for Integrative Sciences and Engineering, 117597 Singapore*

Email: zhangg@ime.a-star.edu.sg

**Abstract**

The thermal conductivity of silicon nanowires (SiNWs) is investigated by molecular dynamics (MD) simulation. It is found that the thermal conductivity of SiNWs can be reduced exponentially by isotopic defects at room temperature. The thermal conductivity reaches the minimum, which is about 27% of that of pure $^{28}$Si NW, when doped with fifty percent isotope atoms. The thermal conductivity of isotopic-superlattice structured SiNWs depends clearly on the period of superlattice. At a critical period of *1.09* nm, the thermal conductivity is only 25% of the value of pure Si NW. An anomalous enhancement of thermal conductivity is observed when the superlattice period is smaller than this critical length. The ultra-low thermal conductivity of superlattice structured SiNWs is explained with phonon spectrum theory.

Silicon nanowires (SiNWs) have attracted a great attention in recent years because of their excellent electrical and mechanical properties [1] and their potential applications in many areas including biosensor.[2] SiNWs are appealing choice because of their ideal interface compatibility with conventional Si-based technology [3, 4]. It is found that the electronic property of SiNWs depends on the surface and growth direction significantly. The band gap is found to be decreased with increasing diameter. [5, 6] These special properties are partially due to the quantum confinement effects in nano materials.

On the other hand, like some other nano materials, the silicon nanowires might be used as thermoelectric materials.[5, 7- 11] The performance of thermoelectric materials depends on the figure of merit ***ZT*** [7]: $ZT = \sigma T S^2 / \kappa$, where *S, T, σ,* and $\kappa$ are the Seebeck coefficient, absolute temperature, electronic conductivity and total thermal conductivity, respectively. ***ZT*** can be increased by increasing *S* or *σ*, or decreasing $\kappa$. Unfortunately, in conventional materials, it is difficult to improve ***ZT***. First, simple increase *S* for general materials will lead to a simultaneous decrease in *σ*. [12] Also, an increase in *σ* leads to a comparable increase in the electronic contribution to $\kappa$. [5, 12,] An alternative way to increase *ZT* is to reduce the thermal conductivity without affecting electronic property. [13] Moreover, ultra-low thermal conductivity is also required to prevent the back-flow of heat from hot end to cool end. Therefore, reduction of thermal conductivity is crucial in thermoelectric application.

Compared with the study of electronic and mechanical properties, much less has done for thermal property of nano materials. [14-21] The thermal properties of nano materials are much different from that of bulk materials. Due to the high surface to volume ratio, the boundary inelastic scatterings affect thermal conductivity significantly. It is shown that the thermal conductivity of SiNWs is about two orders of magnitude smaller than that of bulk crystals. [18, 19] The low thermal conductivity of SiNWs is of particular interest for thermoelectric application.

The thermal conductivity of SiNWs is lower than that of the bulk silicon, whereas it is still larger than the reported ultralow thermal conductivity (0.05 W/m-K) found in layered materials [22]. So it is indispensable to reduce the thermal conductivity of SiNW further in order to achieve high thermoelectric performance. Isotope doping is an efficient way to reduce the thermal conductivity. The isotope effects on thermal conductivity of carbon nanotubes have been investigated by MD simulations [16] and it is found that the thermal conductivity of carbon nanotubes can be reduced more than 50% with isotope impurity. This large reduction by isotopic doping effect has been confirmed experimentally. [17]

In this letter, we shall study the reduction of thermal conductivity of SiNWs with two isotope doping methods. The first one is to dope SiNWs with isotope impurity randomly. The second one is to build isotopic-superlattice structured SiNWs. The total thermal conductivity contains contributions from electrons and phonons, however phonons dominate the heat transport in semiconductors. We thus focus on the isotope effect on phonons transport.

We study the thermal conductivity of SiNWs along (100) direction with cross sections of *3×3* unit cells (lattice constant is *0.543* nm) and ten unit cells in the longitudinal direction which corresponds to cross section area of *2.65* nm$^2$, and length of **5.43** nm. The total number of atoms in the simulation is *720.*

In our simulations, fixed boundary conditions on the outer surface of NW are used and the non-equilibrium molecular dynamics (NEMD) method is adapted. To derive the force term, we use Stillinger-Weber (SW) potential for Si [23], which includes both two-body and three-body potential terms. The two-body potential expression contains a short-distance repulsive term and a long-distance attractive term, and the form is:

$$v_2(r_{ij}) = A\left[B\left(\frac{r_{ij}}{\sigma}\right) - 1\right] \exp\left[\left(\frac{r_{ij} - r_{cut}}{\sigma}\right)^{-1}\right], \quad (1)$$

where $r_{cut}$ is the potential cutoff distance above which no interaction occurs, A, B and $\sigma$ are parameters, $r_{ij}$ is the inter-atomic distance. Since the silicon crystal structure is diamond like, a three-body potential $h_{ijk}$ is introduced to stabilize the correct angular configuration:

$$h_{ijk}(r_{ij}, r_{ik}, \theta_{ijk}) = \lambda \exp[\gamma\sigma[(r_{ij} - r_{cut})^{-1} + (r_{ik} - r_{cut})^{-1}]] \times \left(\cos\theta_{ijk} + \frac{1}{3}\right)^2, \quad (2)$$

where $\theta_{ijk}$ is the angle between $r_{ij}$ and $r_{ik}$, and $\lambda$ and $\gamma$ are parameters. The SW potential has been used widely to study the thermal property of SiNWs and silicon bulk material [18, 24, 25] for its best fit for experimental results on the thermal expansion coefficients.

Generally, in MD simulation, the temperature, $T_{MD}$, is calculated from the kinetic energy of atoms according to the Boltzmann distribution:

$$\langle E \rangle = \sum_1^N \frac{1}{2} m v_i^2 = \frac{3}{2} N k_B T_{MD}, \quad (3)$$

where $\langle E \rangle$ is the mean kinetic energy, $v_i$ the velocity of atom, **m** the atomic mass, **N** the number of particles in the system, $k_B$ the Boltzmann constant. However, this equation is valid only at very high temperature ($T>>T_D$, $T_D$ is the Debye temperature). In our case the average temperature (room temperature) of the system is lower than the Debye temperature ($T_D=645\ K$ of Si), it is necessary to apply a quantum correction to both the MD temperature and the thermal conductivity. Here the average system energy is twice the average kinetic energy based on the equipartition theorem. Then there is an equality of the system energies written in the mechanical and phonon pictures:

$$3Nk_BT_{MD} = \int_0^{\omega_D} D(\omega)n(\omega,T)\hbar\omega d\omega \tag{4}$$

where $D(\omega)$ is the phonon density of states, $n(\omega,T)=1/\left(e^{\hbar\omega/k_BT}-1\right)$ is the phonon average occupation number, and $\omega$ is the phonon frequency, and $\omega_D$ is the Debye frequency. The real temperature $T$ can be deduced from the MD temperature $T_{MD}$ by this relation. [24, 25]

In order to establish a temperature gradient along the longitudinal direction, the two end units of SiNWs are put into heat bathes with temperature $T_L$ and $T_R$ for the left and right end, respectively. Nosé-Hoover heat bathes [26] are used. To ensure our results are independent of heat bath, Langevin heat bathes [27] are also used. Both types of heat baths give rise to the same results. The thermal conductivity is calculated from the Fourier law: $\kappa = -J_l/\nabla T$, where the local heat flux $J_l$ along longitude direction is defined as the energy transported along the NW in unit time through the unit cross-section area, and $\nabla T$ is the temperature gradient. The expression of flux is defined as

$$J_l(t) = \sum_i v_{i,l}\varepsilon_i + \frac{1}{2}\sum_{ij\ i\neq j} r_{ij,l}\left(\vec{F}_{ij}\cdot\vec{v}_i\right) + \sum_{ijk} r_{ij,l}\left(\vec{F}_j(ijk)\cdot\vec{v}_j\right) \tag{5}$$

where $\varepsilon_i$ is local site energy, $\vec{F}_{ij}$ is two-body force and $\vec{F}_j(ijk)$ is three-body force (details see [28]).

Simulations are performed long enough (> $2 \times 10^7$ time steps) such that the system reaches a stationary state where the local heat flux is constant along NW. All results given in this letter are obtained by averaging about $8 \times 10^7$ time steps. A time step is set as *0.8* fs. Using the longitudinal phonon group velocity of Si bulk in (100) direction ($8.3 \times 10^3$ m/s [29]), and the nanowire length (5.43nm), the acoustic wave (phonon) will take $t_0$ ($6.5 \times 10^2$ fs) to propagate from one end to the other through the nanowire. Therefore, within $1.6 \times 10^7$ fs, a heat pulse will go through the system for about $2.5 \times 10^4$ times which is long enough to establish stationary temperature distribution in the wire.

In silicon isotopes, $^{28}$Si is with the highest natural abundance (*92%*), then $^{29}$Si and $^{30}$Si with *5%* and *3%* respectively. [30] Besides these three, $^{42}$Si and $^{43}$Si are also observed. [31] The nuclide $^{42}$Si has become the focus of particular interest as it is a unique opportunity to study the nuclear shell effects. Many experimental and theoretical studies [32-36] have been conducted on the isotopic effect on thermal conductivity of bulk Si. However, the results have been controversial. The first study of isotopic pure bulk $^{28}$Si reported that the thermal conductivity at room temperature was enhanced by 60% relative to natural Si [32], and calculations found agreement between theory and experiment [33]. But the authors of Ref. 32 later retracted that conclusion and reported a change of only 10% at room temperature [34]. Other experimental studies of $^{28}$Si single crystals obtained similar values [35], and some theories predict a 10%-20% enhancement [36]. In this paper, the effect of doping $^{29}$Si and $^{42}$Si to $^{28}$Si NWs are studied, because it is convenient to dope $^{28}$Si with $^{29}$Si. The mass difference between $^{42}$Si and $^{28}$Si is large, which provides a good chance to study the mass influence on thermal conductivities.

Firstly, we study the effect of randomly distributed isotopic atoms on the thermal conductivity of $^{28}$Si NWs. In our simulations, in order to reduce the fluctuation, the results are averaged over five realizations.

As shown in Figure 1, the value of thermal conductivity is *1.49* W/m-K for pure $^{28}$Si NW, which agrees with previous simulation and experimental results.[18, 19] The thermal conductivity of SiNW is about two orders of magnitude smaller than that of bulk silicon crystal. This is due to the following facts. Firstly, the low frequency phonons, whose wave lengths are longer than the length of nanowire, cannot survive in nanowire. Therefore, the low frequency contribution to thermal conductivity, which is very substantial and significant, is largely reduced. Secondly, because of the large surface to volume ratio, the boundary scattering in quasi-1D structure is also significant. The curve of thermal conductivity decreases first to reach a minimum and then increases as the percentage of isotope impurity atoms changes from zero to hundred. Interestingly, the conduction curves follow exponential law. So, at low isotopic percentage, the small ratio of impurity atoms can induce large reduction on conductivity. For example, in the case of $^{28}$Si$_{0.98}$$^{42}$Si$_{0.02}$, namely, two percent $^{42}$Si, its thermal conductivity is only *65%* of that of pure $^{28}$Si NW. Contrast to the high sensitivity at the two ends, the thermal conductivity versus isotopic concentration curves are almost flat at the center part as show in Figure 1, where the value of thermal conductivity is only 27% ($^{42}$Si doping) or 77% ($^{29}$Si doping) of that of pure $^{28}$Si NW.

It is worth pointing out that $^{42}$Si is not as stable as $^{29}$Si and $^{30}$Si. We study the isotopic doping with $^{42}$Si because the large mass difference between $^{42}$Si and $^{28}$Si thus we can explore the mass influence on isotopic effect. To compare with the experimental results



in bulk Si, we also calculate the thermal conductivity with natural isotopic abundance, 5% $^{29}$Si and 3% $^{30}$Si. The calculated thermal conductivity is 1.28 W/m-K, around 86% of pure $^{28}$Si NW, which is close to the experimental results in bulk Si [34, 35].

Next we study the thermal conductivity of $^{28}$Si NW doped with two different isotope atoms ($^{29}$Si and $^{42}$Si) at the same time. The sum of isotope atom's percentages is fixed at 50%, and we change the ratio of numbers of the two isotope atoms. The results are shown in Table 1. And the thermal conductivity is *1.17* W/m-K for $^{28}$Si$_{0.5}$$^{29}$Si$_{0.5}$ NW, which is larger than that of $^{28}$Si$_{0.5}$$^{42}$Si$_{0.5}$, *0.40* W/m-K. It is obvious that the heavier isotopic atoms can induce larger reduction on thermal conductivity than that from lighter isotopic atoms.

Besides isotopic effect, the fact that the thermal conductivities of both bulk and nano scale materials are greatly influenced by vacancies has been confirmed experimentally and theoretically. [14, 37-39] The reduction mainly comes from the high frequency phonons, which are sensitive to the vacancy [37]. With MD simulations, it is found that the thermal conductivities of diamond and carbon nanotubes decrease as the vacancy concentration increase with an inverse power law relation, $\kappa \propto x^{-\alpha}$, where the exponent α is 0.79 for carbon nanotube and 0.69 for diamond crystal. [14, 37] The MD calculation results show that vacancies in nanotubes are not much more influential than in 3-D diamond. While in this letter, we find that the thermal conductivity of SiNW decreases exponentially with the increase of isotope concentration. The vacancy effect on thermal conductivity of SiNW needs to be studied further.

Some experimental and theoretical works [40-44] have been carried out to study the effects of interface and superlattice period on thermal conductivity of various kinds of superlattice structures. Results show that superlattice structures are an efficient way to get



ultra low thermal conductivity. However for structures built from different crystal materials, the relatively high interface energy will limit the stability of these structures. The isotopic-superlattice (IS) is a good structure to reduce the thermal conductivity without destroying the stability. On the other hand, most of the studies, especially experimental ones, have focused on the structures of the superlattice period longer than the phonon mean free path.[40-43] Here we study the thermal conductivity of IS structured NWs which consists of alternating $^{28}$Si/$^{42}$Si or $^{28}$Si/$^{29}$Si layers along the longitudinal direction and with ultra-short superlattice period lengths (see Fig. 2). We select the SiNW longitudinal axis as the *x* axis, atoms in the same layer means they have the same *x* coordinate. The NW length is fixed at 5.43 nm (10 unit cells) and with cross sections of *3×3* unit cells. Figure 3 shows the thermal conductivity of the IS SiNWs versus the superlattice period length. As expected, the thermal conductivities of $^{28}$Si/$^{42}$Si are much lower than those of $^{28}$Si/$^{29}$Si, so the mass effect on reduction of thermal conductivity is also obvious in the IS structures. Although there is difference in absolute values of thermal conductivity between $^{28}$Si/$^{42}$Si and $^{28}$Si/$^{29}$Si structures, the dependence of thermal conductivity on superlattice period length is similar. The thermal conductivity decreases as the period length decreases from *4.34* nm (32 layers) to *1.09* nm (8 layers), which means the number of interface increases. The thermal conductivity reaches a minimum when the period length is *1.09* nm. This is consistent with the fact that increasing number of interface for fixed length will enhance the interface scattering which gives rise to the reduction of the thermal conductivity. As the superlattice period decreases further, interesting phenomena appears. With the period length smaller than *1.09* nm, the thermal conductivity increases rapidly as the period length decreases. At the



smallest period length of *0.27* nm (2 layers, 1 layer $^{28}$Si plus 1 layer $^{42}$Si or $^{29}$Si), the thermal conductivity of IS SiNW increases to *0.85* W/m-K ($^{28}$Si/$^{42}$Si) or *1.38* W/m-K ($^{28}$Si/$^{29}$Si).

This anomalous increase in thermal conductivity can be understood qualitatively from the dominant phonon wavelength theory. At room temperature, the dominant phonon wavelength is about 1~2 nm in most materials [45], which is very close to the critical period length of *1.09* nm found in SiNWs in this letter. So it can be concluded that the thermal conductivity will increase significant after decreasing the superlattice period length to be smaller than the dominant phonon wavelength. This result coincides with Simkin's result on harmonic model. [46]

To get a better understanding of the underlying physics of thermal conductivity reduction in superlattice NWs and the anomalous increase phenomena, we calculate the phonon density of states (DOS) of $^{28}$Si layer and $^{42}$Si layer in two IS structured SiNWs: 2.17 nm-period NW and 0.27 nm-period NW as shown in Figure4. When period length is *2.17* nm (16 layers), there is an obvious mismatch in the DOS spectra, both at low frequency band and high frequency band, results in very low thermal conductivity (Fig.3). On the contrary, the DOS spectra overlap perfectly for IS structured SiNW whose period length is *0.27* nm (2 layers). The good match in the DOS (Fig.4 (b)) comes from the collective vibrations of different mass layers which is harder to be built in longer superlattice period structures. The match/mismatch of the DOS spectra between the different mass layers controls the heat current. When the two spectra largely overlap with each other, the heat current can easily go through the system, thus results in high



conductivity. On the other hand, when the spectra mismatch, the heat current is difficult to go through the structure and thus leads to very low thermal conductivity.

To summarize, we have obtained ultra-low thermal conductivity of SiNWs by isotopic doping. Two kinds of methods have been used to reduce thermal conductivities. The first method is random doping NW structure. The random doping can lead to a large scale decrease of thermal conductivity of SiNW, even with only two percent isotopic atom. And the second method is isotopic doping in superlattice structure. The superlattice structure can reduce the conductivity significantly because the mismatch in DOS spectra of different mass layers. It is interesting that the thermal conductivity increases anomalously when the superlattice period length is smaller than a critical value (*1.09* nm in SiNWs). It is shown that there are collective vibrations of different mass layers in IS structured SiNWs when superlattice period is smaller than this critical value. The collective vibrations result in large overlap in DOS spectra. The mass effect on conductivity is obvious. The heavier isotope atoms ($^{42}$Si) can decrease conductivity much more than the lighter ones ($^{29}$Si). This is consistent with the idea that to increase *ZT*, the only way to reduce $\kappa$ without affecting *S* and $\sigma$ is to use heavier atoms.[13] The remarkable isotopic effect observed in this work provides an efficient approach to decrease thermal conductivity of SiNW. This is of great benefit to improve the thermoelectric performance as low thermal conductivity also can prevent the back-flow of heat. These improvements have raised the exciting prospect that SiNWs can be applied as novel nano scale thermoelectric materials.

The work is supported by the Temasek Young Investigator Award from DSTA



Singapore under Project Agreement POD0410553 and an Academic Research Fund grant, R-144-000-203-112, from Ministry of Education of Republic of Singapore.

[45] G. Chen, D. Borca-Tasciuc, and R.G. Yang, "Nanoscale Heat Transfer" in "Encyclopedia of Nanoscience and Nanotechnology", eds. H.S. Nalwa, American Scientific Publishers (2004), Vol. 7, pp. 429-459.

[46] M.V. Simkin and G. D. Mahan, Phys. Rev. Lett. **84**, 927 (2000).16

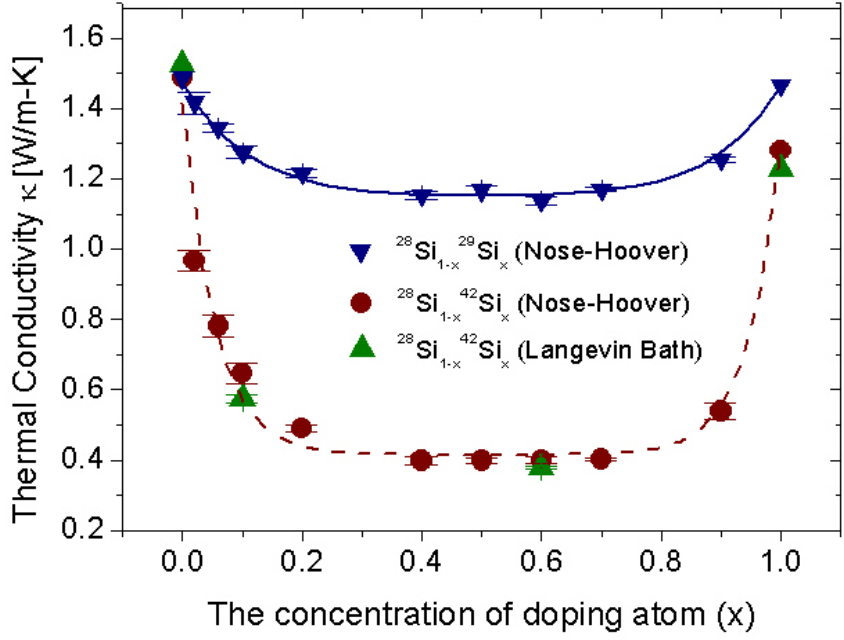

FIG. 1. Thermal conductivity of SiNWs versus the percentage of randomly doping isotope atoms ($^{42}$Si and $^{29}$Si) at 300K. SiNWs are along (100) direction with cross sections of **3×3** unit cells (lattice constant is **0.543** nm) and ten unit cells in the longitudinal direction. The results by Nose-Hoover method coincide with those by Langevin methods indicating that our results are independent of the heat bath used. The solid/dash curve is the best fitting to the formula $\kappa = A_1 e^{-x/B} + A_2 e^{-(1-x)/B} + C$, where $A_1 = 0.33, A_2 = 0.32, B = 0.11,$ and $C = 1.2$ for $^{42}$Si, and $A_1 = 1.00, A_2 = 0.88, B = 0.054,$ and $C = 0.42$ for $^{42}$Si.



TABLE 1. Thermal conductivity of SiNWs with two kinds of randomly doping isotope atoms ($^{42}$Si and $^{29}$Si).

|  | $^{28}$Si | $^{28}$Si$_{0.5}$$^{29}$Si$_{0.5}$ | $^{28}$Si$_{0.5}$$^{29}$Si$_{0.4}$$^{42}$Si$_{0.1}$ | $^{28}$Si$_{0.5}$$^{29}$Si$_{0.3}$$^{42}$Si$_{0.2}$ |
|---|---|---|---|---|
| κ (W/m-K) | 1.49 | 1.17 | 0.64 | 0.48 |
|  |  | $^{28}$Si$_{0.5}$$^{29}$Si$_{0.2}$$^{42}$Si$_{0.3}$ | $^{28}$Si$_{0.5}$$^{29}$Si$_{0.1}$$^{42}$Si$_{0.4}$ | $^{28}$Si$_{0.5}$$^{42}$Si$_{0.5}$ |
| κ (W/m-K) |  | 0.46 | 0.40 | 0.40 |

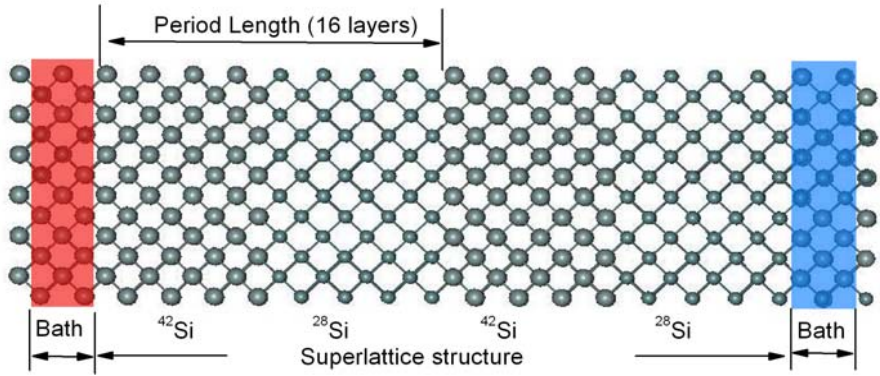

FIG. 2. Schematic picture of the IS structured SiNW. The period length is *2.17* nm (16 layers) here. Atoms with different mass are denoted by the different size.

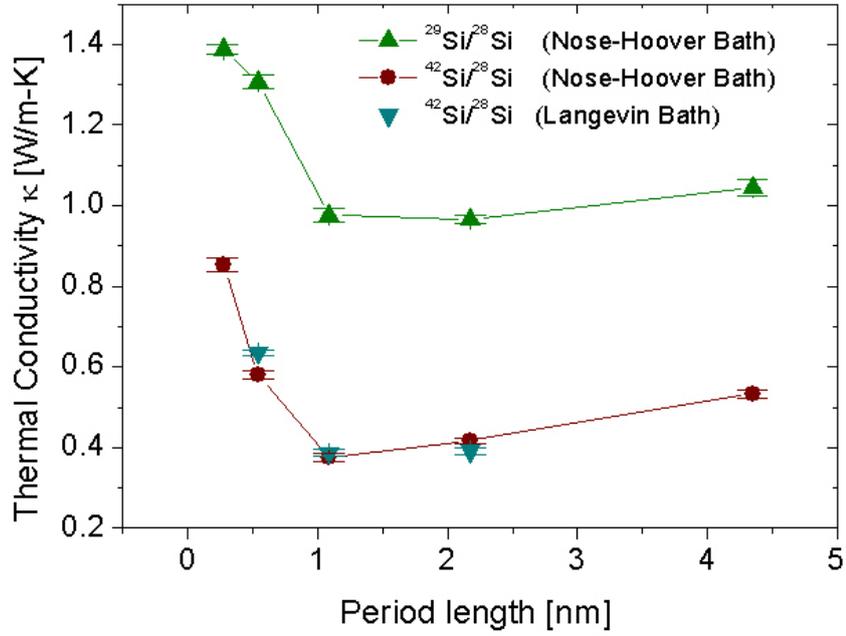

FIG. 3. Thermal conductivity of the superlattice SiNWs versus the period length at 300 K. SiNWs are along (100) direction with cross sections of *3×3* unit cells (lattice constant is *0.543* nm) and ten unit cells in the longitudinal direction. The results by Nose-Hoover method coincide with those by Langevin methods indicating that our results are independent of the heat bath used.

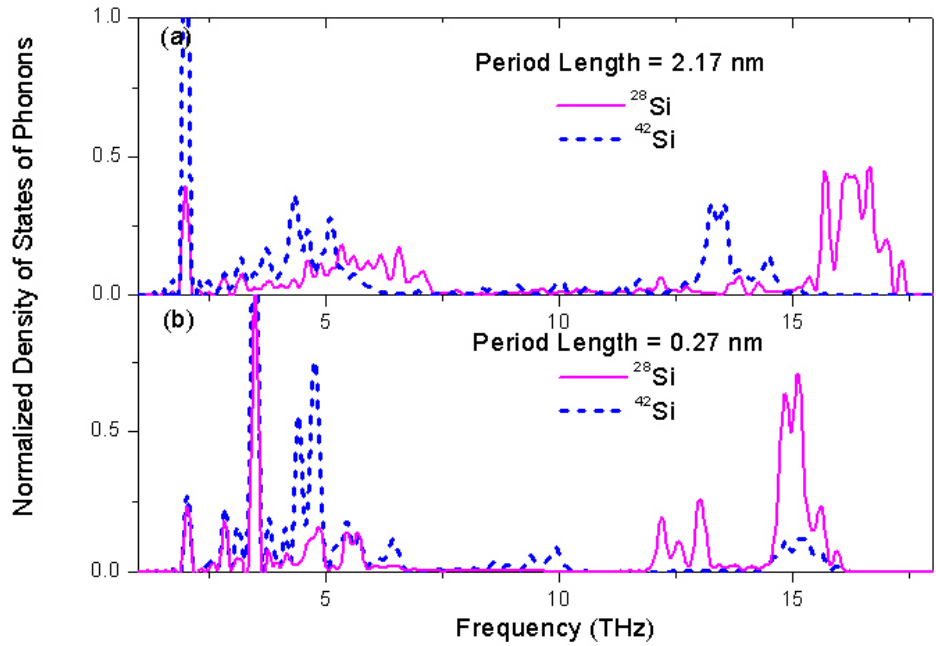

FIG. 4. The average normalized density of states of phonons of different mass atoms ($^{28}$Si or $^{42}$Si) along the longitude direction for superlattice SiNW structures. (a) The period length of superlattice NW is 2.17 nm (16 layers). This structure has low conductivity, *0.42* W/m-K. (b) The period length of superlattice NW is 0.27 nm (2 layers). This structure has high conductivity, *0.85* W/m-K.